# Activism by the AI Community

Analysing Recent Achievements and Future Prospects

Haydn Belfield
Centre for the Study of Existential Risk
University of Cambridge
Cambridge, UK
hb492@cam.ac.uk

## ABSTRACT

The artificial intelligence (AI) community has recently engaged in activism in relation to their employers, other members of the community, and their governments in order to shape the societal and ethical implications of AI. It has achieved some notable successes, but prospects for further political organising and activism are uncertain. We survey activism by the AI community over the last six years; apply two analytical frameworks drawing upon the literature on epistemic communities, and worker organising and bargaining; and explore what they imply for the future prospects of the AI community. Success thus far has hinged on a coherent shared culture, and high bargaining power due to the high demand for a limited supply of AI 'talent'. Both are crucial to the future of AI activism and worthy of sustained attention.

## CCS CONCEPTS

• Social and professional topics~Computing profession • Social and professional topics~Computing and business • Social and professional topics~Government technology policy

## KEYWORDS

Activism, organising, ethics, epistemic community, lethal autonomous weapons, bargaining, talent supply.

## 1  Introduction

"The best thing AI researchers can do is vote with their feet, not work with companies that have outcomes you don't agree with. There aren't enough researchers to go around, and attracting enough talent is important, so actually researchers individually have a lot of power. Through soft influence, you can influence a lot."
–Demis Hassabis, DeepMind CEO [49]

The development and deployment of AI is likely to have important consequences for the global economy, society, and politics. AI deployment is already influencing markets and the economy, justice and distributive decisions by governments, and our elections [32, 63]. In coming years we may see the use of AI systems with novel capabilities in military domains [47]. Increasing use of increasingly powerful yet brittle systems carries accident risks, misuse risks, and structural risks [65] – risks to human security, and risks of societal inequality and discrimination [6, 37].

The AI community has responded to these urgent issues by engaging in activism in order to promote the positive (in their view) societal and ethical effects of AI, and decrease the negative effects. 'Activism' includes a broad range of different kinds of social and political campaigning, organising and advocacy. This encompasses: issue-framing; agenda-setting; standard setting; private discussions with decision-makers; public campaigning in traditional and social media; establishing new fields and organisations; submissions to governmental inquiries; and classic labour tactics of boycotts and strikes.

The 'AI community' includes researchers, research engineers, faculty, graduate students, NGO workers, campaigners and some technology workers more generally – those who would self-describe as working 'on', 'with' and 'in' AI and those analysing or campaigning on the effects of AI. This paper focuses especially on the AI community within corporate and academic labs in the US and Europe.

This activism has had some notable consequences so far: informing international negotiations, changing corporate strategy, and spurring the growth of research fields (see next section). Such activism may shape the manner and extent to which AI is militarised, and how AI companies address ethics and safety concerns. The AI community is an important autonomous actor with a distinctive set of viewpoints and interests. It needs to be accounted for in strategic or academic analysis and negotiated with by other actors. AI community activism may profoundly shape the development and deployment of this important set of technologies - and therefore shape our global economy, society and politics.

This activism is therefore an important phenomenon in need of theoretical analysis. To date this analysis has been limited, with much of the discussion in 'grey literature' or in the media [60, 20, 55, 11]. We can identify (at least) two important questions for researchers:
- How can we explain the activism by the AI community of recent years?
- What are the future prospects of activism by the AI community?

In this paper, we overview some examples of recent activism by the AI community. We then apply two different analytical frameworks: epistemic communities and worker organising. We end by discussing some key common factors, and identifying research questions that could clarify the future prospects of activism by the AI community.

## 2 Recent examples: 2013-2019

This section is a brief overview of some examples of activism by the AI community over the last six years. The purpose of this section is to motivate the discussion by outlining some concrete examples to ground the later analytical discussion. See AI NOW's 'Year in Review' reports [9, 10, 11] for a more comprehensive descriptive treatment.

The last six years have seen a number of high-profile political actions taken by the AI community. We discuss three main examples: lethal autonomous weapons systems (LAWS), ethics and safety, and employee organising. These three do not cover the entire range of activism, but are intended to be broadly illustrative. Under these headings we proceed in a roughly chronological manner. We begin in 2013 with the launch of the Campaign to Stop Killer Robots, a key moment of epistemic community formation and a focus for worker organizing[1].

### 2.1 Lethal Autonomous Weapons Systems (LAWS)

LAWS are weapons systems that can autonomously select and kill targets [35]. The AI community has been deeply involved with the debate over an international ban on LAWS [51]. Not only is the AI community a relevant expert community for advising on this debate, it is also being directly asked to work on the research and development (R&D) of LAWS [59].

In early 2013 the Campaign to Stop Killer Robots was formally launched to promote an international ban on the development and use of LAWS. In late 2013 the Convention on Certain Conventional Weapons (CCW) agreed to begin considering LAWS. The CCW was the forum for negotiations over banning cluster munitions, landmines and blinding laser weapons. The AI community played key roles in the adoption of LAWS as an issue by arms control NGOs, the establishment of the UN process, and the ongoing work of the CCW [4].

The Campaign includes many members of the AI community. Their activism ranges from personal discussions with diplomats at the CCW to mass media 'viral videos' [7]. A key tactic has been the organisation of mass Open Letters. In July 2015 an Open Letter on LAWS expressed community concern. It has been signed to date by 4,500 AI and robotics researchers [22]. This was followed by another Open Letter on LAWS [24], when the Meeting of Experts changed to a Group of Governmental Experts (GGE) at the CCW.

This effort has also included intra-community organising. One example is the April 2018 Korea Advanced Institute of Science and Technology (KAIST) boycott. The 50 signatories committed to boycott all collaborations with any part of KAIST, due to concerns that a KAIST centre had a LAWS collaboration with Hanwha Systems, a leading South Korean arms company. KAIST soon clarified that they would not work on LAWS [25].

Activism has also included extensive and intense negotiations within technology companies, discussed below.

### 2.2 Ethics and safety

There has been sustained activism from the AI community to emphasise that AI should be developed and deployed in a safe and beneficial manner. This has involved Open Letters, AI principles, the establishment of new centres, and influencing governments.

The Puerto Rico Conference in January 2015 was a landmark event to promote the beneficial and safe development of AI. It led to an Open Letter signed by over 8,000 people calling for the safe and beneficial development of AI, and a research agenda to that end [21].

The Asilomar Conference in January 2017 led to the Asilomar AI Principles, signed by several thousand AI researchers [23]. Over a dozen sets of principles from a range of groups followed [61].

The AI community has established several research groups to understand and shape the societal impact of AI. AI conferences have also expanded their work to consider the impact of AI. New groups include:
- Fairness, Accountability, and Transparency in Machine Learning (FAT ML) (December 2014)
- OpenAI (December 2015)[2]
- Centre for Human-Compatible AI (August 2016)
- Leverhulme Centre for the Future of Intelligence (October 2016)[3]
- Algorithmic Justice League (November 2016)
- DeepMind Ethics and Society (October 2017)
- AI Now Institute (November 2017)
- UK Government's Centre for Data Ethics and Innovation (November 2017)[4]

Especially notable is the Partnership on Artificial Intelligence to Benefit People and Society (September 2016). The Partnership brings together over 90 companies and non-profits, and is exploring best-practice recommendations for the community as a whole [46].

---
[1] The campaign's roots stretch back at least to 2004 [8].
[2] OpenAI was a non-profit and is now a company with a non-profit mission.
[3] The author is an Associate Fellow of the Centre.

[4] The Centre was proposed and advocated for by the AI community, and analyses AI's societal and ethical implications.

The AI community has also contributed to over 30 national and international AI strategies [13]. The High-Level Expert Group on Artificial Intelligence to the European Commission (HLEG-AI) is made up of 52 experts from academia, civil society, and industry. Their work led to Ethics Guidelines, and policy and investment recommendations, for trustworthy AI [33]. EU Commissioner Margrethe Vestager has strongly indicated that forthcoming EU regulation on AI will be based on this work by the AI community.

## 2.3 Organising

2018 saw an upturn in political organising, especially within large technology companies: the 'tech resistance'.

The most prominent example of this activism has been Google. Google's involvement with Department of Defense's (DoD) Project Maven was revealed in March 2018. 3,000 employees signed an Open Letter opposing it in April [28]. In June, Google announced it would not renew the Project Maven contract and released its AI Principles. In October, Google also dropped out of the DoD's JEDI cloud bidding process.

However, in August, Google's secret 'Project Dragonfly' (a censored Chinese search engine) was revealed. This was opposed by another Open Letter. Two anonymous employees wrote in an email circulating the Dragonfly letter: "Individual employees organizing against the latest dubious project cannot be our only safeguard against unethical decisions. This amounts to unsustainable ethics whack-a-mole" [29].

In November, 20,000 employees and contractors took part in a one-day strike, or 'Google walkout', protesting sexual harassment and misconduct. This contributed to the end of 'forced arbitration' for full time workers [30].

Not all organising, however, was so clearly successful. Ongoing debates include corporate partnerships with Immigration and Customs Enforcement (ICE) and Customs and Border Protection (CBP). In June, perhaps as a result of the Trump administration's family separation policy [10], several groups released open letters [42, 2, 52].

## 3 Analytical frameworks: two lenses

We reviewed three examples of activism over the last six years, including some key successes. We now draw on two different analytical frameworks: epistemic communities, and worker organising. Both of these can provide insights into why this activism is occurring, how these successes have been achieved, and what its future prospects are.

We address each framework in turn. For each framework, we briefly describe it and explain its relevance to our discussion. We then discuss what factors are seen as predictive of success for the relevant group within the framework, and ask to what extent those factors apply to our case – and will continue to apply over the next few years.

## 3.1 Epistemic communities

An epistemic community is a network of knowledge-based experts: "professionals with recognised expertise and competence in a particular domain and an authoritative claim to policy-relevant knowledge within that domain" [31]. Crucially, they also share causal and principled beliefs about their domain, notions of validity, and a common policy project [1].

Historical examples include nuclear physicists, chemists, biologists, and climate scientists [31]. Epistemic communities have been identified as playing key roles in several policy debates, such as the role of the nuclear weapons scientists and strategists in the 1972 Anti-Ballistic Missile Treaty [1].

This framework is relevant to the AI community. They are technical experts that have a clearly recognised and valid claim to authority and expertise in the domain of AI. There is a strong set of shared causal beliefs and notions of validity, and common policy projects.

Scepticism has been expressed about the coherence of the epistemic community on military AI, especially about whether it is presently sufficient to play a similar role to the ABM community [39, 47]. It clearly has not yet been sufficient to pass an international ban or block military AI R&D efforts. However, it has put LAWS firmly on the international agenda, and the Campaign's prospects look similar to the landmines and cluster munitions campaigns at similar stages. For now, it has temporarily paused, and perhaps lastingly complicated, the relationship between the US military and US tech firms (most importantly Google). Also, we are considering activism on a wider set of topics than just LAWS. The AI community's policy projects also include emphasising ethics, societal benefit and safety, and opposing particular choices by company management.

In general, when are epistemic communities more likely to be persuasive? Cross [12] identifies five key factors: 1) the issue is uncertain and salient; 2) the community has access to, and understanding of, decision-makers and other actors; 3) there is 'policy field coherence', 4) they seek to influence an early or technocratic phase in the policy process; and 5) they are seen as credible and have a more cohesive, certain coalition than their competitors.

We suggest these factors currently apply to the AI community, and are likely to largely continue to apply.
1) The ethical, military and societal implications of AI is a complex and new set of issues. It has become politically salient over the last few years.

It is likely to remain quite uncertain. The extent to which there is a sense of perceived crisis may fade over time as the issue becomes less new and begins to be untangled. However, it is likely to remain technically complex, as well as politically salient, as AI continues to impact our economy and society.
2) The AI community has access to top decision-makers, both corporate and political. They have also been able to anticipate other actor's preferences and actions, such as anticipating those of the different national delegations during the international LAWS negotiations.

This high-level access is likely to continue – policy-maker interest sees no sign of slacking. It is open whether they will continue to be anticipatory.

3) 'Policy field coherence' refers to there being respected quantitative data, the issue involving technical systems, and the norms and goals of the community being compatible with existing institutional norms. The issues for the AI community have involved the interaction of human-made technical systems with social systems, instead of just involving social systems, and there has been respected quantitative data on, for example, the extent of bias or the frequency of mistakes by ML systems [37].

There will likely continue to be policy field coherence. Indeed the availability of quantitative data is likely to improve over time. However the lack of quantitative data about the humanitarian impact of LAWS (which will continue until if and when they are deployed) may continue to limit the persuasiveness of those pushing for an international ban.

4) The AI community has sought to influence the initial terms of the debate and to focus on subsystem, technocratic phases. For example, activism in 2015 helped shape the terms of the debate around ethical, safe and beneficial AI – and then how that was translated into national ethical principles and strategies.

Over time, the terms of the debates around the development and deployment of AI are likely to become more fixed, and activism will have to seek to influence later stages in decision-making processes. The societal implications of AI may also become more pronounced, and/or entangled with broader political beliefs. However, continued technical breakthroughs may create continued capability jumps (such as language models and fake news generation), routinely posing new policy problems.

5) The AI community has been quite coherent and certain of its aims in some respects: pushing for serious consideration of the military and societal impacts of AI by companies and governments. It seems more cohesive than those pushing to deproblematise the issue. It also shares a high level of professional norms and status.

It is unclear whether they will continue to be more coherent and certain than competing networks, especially in the debate over the acceptability of working on military AI, including LAWS. The network pushing for an accommodation with militaries may prove more cohesive and certain. Community credibility seems unlikely to radically change. Marginal changes could occur, for example through the further democratisation of machine learning (ML) through MOOCs and other online platforms.

Overall then, the AI community has achieved some successes as an epistemic community as the scope conditions (1), political opportunity structure (2) and policy field coherence (3) have been favourable, and the AI community has sought to influence early and technocratic phases in the policy process (4) and built fairly cohesive coalitions (5).

Factors likely to continue include high-level access being good; being able to deal with technocratic aspects of decision-making; sharing a high status; and being able to draw on evidence on which they are expert. However, over the next few years novelty will decrease, the terms of debate will become more set, and the entire issue-area may become more politicised. The window of influence may be shrinking, but it is unlikely to close, as the issue-area will remain uncertain and complex. A key question for further research is whether the AI community will continue to be cohesive and certain in its aims.

## 3.2 Worker organizing and bargaining

Freeman and Medoff [19] influentially distinguish between the 'two faces' of organised labour. The first, 'monopoly face' is that of union monopoly power used to raise members' wages. Now, however, the emphasis is typically more on offsetting informational and power asymmetries between workers and management [5].

More relevant to our focus, however, is their second, "collective voice/institutional response" face. This builds on Hirschman's [34] conception of 'voice' as the ability to "change, rather than escape from [i.e. exit], an objectionable state of affairs."

Worker organising and bargaining means their ability "to 'voice' their concerns and demands rather than immediately 'exit'—that is, quit the job", "to voice complaints and see them addressed through collective bargaining" [19]. Worker or employee voice – whether formally through a union or not – has become a key topic for disciplines such as labour economics, industrial relations and organisational behaviour [62].

It need not refer just to improving working conditions, but also to encouraging one's firm to advocate for public policies. For example, US corporate engagement on LGBT rights has been largely driven by employee organizations in highly-educated workforces advocating for management to take public stands advocating for LGBT rights [40].

Key to this framework is the ability of different actors to have an agreement on their terms, that is the 'bargaining power' of workers and management, based on the "ability to impose costs on the other side for failing to agree and to avoid or absorb its own costs from failing to agree" [14].

This framework is relevant to the AI community. Almost all members of the AI community are employees, rather than business-owners. Indeed, the AI community is largely located within fairly large corporate and academic labs.

Formal unionisation rates are low in technology companies, though higher in academia [18]. However, it is clear that the AI community has substantial bargaining power, of the type described in our opening quote. This is reflected in high salaries. The New York Times reports that "A.I. specialists with little or no industry experience can make between $300,000 and $500,000 a year in salary and stock. Top names can receive […] millions" [41]. Nevertheless, groups like the Tech Workers Coalition have emphasised the difference in power across the industry – from a researcher with a ML PhD to a gig economy Mechanical Turk contractor. The Coalition's work began in attempting to organise across more of the tech workforce [58]. The organizing examples described above often involved workers adapting classic collective bargaining tactics for their situations.

While this framework is directly relevant to employees, more generally, AI community activism can be framed as a set of strategic interactions between actors, each with different incentives, resources and constraints – where outcomes are the result of bargaining between these actors [38].

In general, when are workers more likely to be persuasive (or successful) with the management of their organisation? Dau-Schmidt & Ellis [14] identify five key economic factors: 1) the nature of the organisation's product and services; 2) the structure of bargaining; 3) the organisation's technology of production; 4) general economic conditions; and 5) the employees' commitment to collective action.

We suggest these factors currently benefit the AI community, and are likely to largely continue to benefit it.

1) Corporate organisations' products and services typically require ongoing maintenance, so management is less able to absorb costs from failing to agree with its employees. Also, large tech companies are generally consumer-facing (B2C) rather than business-facing (B2B) firms, which makes them more susceptible to public – consumer – sentiment. Google is more susceptible than Palantir to consumer sentiment because they rely on the public using their services, rather than Palantir which relies more on large clients. Tech workers have been adept at mobilising media interest and public support. The nature of these products and services is unlikely to change dramatically in the coming years.

2) The structural landscape of bargaining is largely tilted in favour of employers. While many tech workers are well-paid enough to have a decent amount of runway, and are able to 'exit' to similarly well-paid jobs elsewhere, they are hampered by the lack of unionisation. Another issue is the widespread use by tech companies of non-disclosure agreements (NDAs). This tends to limit workers' ability to publicly voice concerns – even after they leave the company.

This is likely to continue, but may diminish marginally if there are explicit offers of financial or legal support from worker groups like the Tech Worker's Coalition or civil society groups like the American Civil Liberties Union (ACLU). However, several centre-left politicians in the US and Europe propose more antitrust scrutiny of tech companies, and to change union regulation to empower them, which would shift the landscape.

3) Organisations employing members of the AI community typically depend on highly-skilled workers. Replacements are hard to obtain, and the organisation cannot continue with just a skeleton crew. This means that the organisations are less able to absorb the costs of failing to agree a deal, improving worker bargaining power.

This is likely to continue, but may change marginally. If these organisations turn from more R&D-style work to more implementation-style work, then workers may not need to be as highly-skilled. This may shift bargaining power.

4) A key question about the general economic conditions is the balance of supply and demand of potential replacement workers. Estimates of the 'talent pool' vary widely, from 22,400-36,500 [26] to 300,000 [54]. The high salaries commanded in the field demonstrate that supply is not meeting demand.

However, various sources of evidence point to an increase in the talent supply in coming years [53]. Several governments have made large commitments of money for education: the UK is supporting 1,000 PhDs [16]. The 2017 CRA Taulbee Survey shows a continued increase in the number of degrees and enrolment, at doctorate, masters and undergraduate level [64]. Note that this does not include quasi-academic routes: MOOCs such as Andrew Ng's or fast.ai, coder bootcamps, and so on.

We should expect an increase in the supply of talent over the coming years (a PhD in ML, for example, typically takes four years to complete in the UK and six years in the USA). Other things being equal, this increase in the number of potential workers would tend to decrease worker power.

However, this assumes that supply increases more than demand. The following years may well see demand for ML talent continue to increase – keeping track with or even exceeding the increase in supply. This demand may increase as AI and robotics becomes more technologically ready or industrialised. If data and computing power continue to grow, more areas may become well-suited to ML approaches, which might also increase demand. The future balance of supply and demand is a key question.

5) Commitment to collective action is a similar factor to epistemic community cohesion. Willingness and ability to (threaten) exit seems fairly high. There have been several high-profile exits over 'employee voice' issues [44, 45, 48]. A small-scale survey of UK tech workers suggests 16% of all people in AI have left their company over the issue of working on products they felt might be harmful for society, compared to 5% of all tech workers [43].

A key question is whether the predicted new 'talent' will be socialised to a similar extent, and through similar means, to existing workers. A larger AI community, that has not all been to the same small academic conferences for several years, may be less cohesive. Conversely, enrolment in university 'ethics in AI' courses may make the next generation more ethically engaged. Also, if the momentum behind the political organising continues, it will increasingly be seen as the norm. Key organisers might receive more training and support. The ad-hoc arrangements that have characterised the political organising of the AI community so far might become more institutionalised.

Overall then, the AI community has achieved some successes as workers organising and bargaining with their employers. This may be attributed to the organisational products and services (1), organisational production technology (3), and the general economic conditions (4) all being favourable – though the structure of bargaining has not been (2) – and the AI community having been fairly committed to collective action (5).

Factors likely to continue include the products and services being consumer-facing and difficult to stockpile; reliance on high-skilled labour; and the unequal bargaining structure. However it is unclear what the balance of talent supply and demand will be, and to what extent the AI community will continue to be committed to collective action. These are key questions for further research.

## 4 Discussion and further research

What are the prospects for the AI community being able to successfully continue its activism over the coming years? Two factors have been identified as key questions for further research: the balance of talent supply and demand and the cohesiveness of the AI community.

We identified a small, narrow, talent pool as a key factor in the bargaining power of the AI community. We need to be able to assess relative supply and demand of AI talent over the coming years. Further quantitative research is needed into, for example, the current size of the talent supply, collated salary information, enrolment rates, or future demand projections for AI talent.

The AI community has several resources at its disposal to maintain and deepen its social, organisational, ethical and political cohesion. The AI community has a strong shared culture, with strong norms of responsibility, 'do-it-yourself' and mutual support. The community publish at and attend the same major conferences (such as AAAI, NeurIPS, ICML and IJCAI), and publish on the same sites such as arXiv and GitHub. Virtually none of these outlets charge for publication or access, and are often maintained by the community. This culture can be seen in the closed-access journal boycott [17] and the name change to NeurIPS [3].

The AI community has made use of institutional enabling structures: the Future of Life Institute as a coordinator of international Open Letters; the Tech Workers Coalition and the Partnership on AI as distributors of best practice; corporate digital tools such as email mailing lists and internal chat rooms; and social media such as Twitter and Medium as a way of communicating demands. These structures could be developed, better funded and institutionalized – and key organisers could receive more training and support. Further research into the AI community's ability and willingness to maintain and deepen its cohesion – and the structures and institutions that support that – is needed.

Further research could also extend our analysis to a broader range of regional and sectoral contexts, to a broader range of actors with influence over companies, and to other analytical frameworks.

## 5 Conclusion

The AI community is acting together – it is organised. It has won some key successes. And yet its future prospects are uncertain. Useful insights can be drawn from the literature on epistemic communities and worker organising. There are good precedents for highly-skilled groups engaging in activism as epistemic communities and valuable employees, thereby influencing policy and corporate outcomes. Indeed, the AI community has already had some clear successes around LAWS, ethics and safety, and employee organising. Whether this will continue is less clear. The future balance of AI talent supply and demand and the future cohesion of the AI community are key questions for further research.

## 6 Appendix: Veto players

Veto players are individual or collective actors whose agreement (by majority rule for collective actors) is required for a change of the status quo [56]. This framework is generally applied to comparative politics, where it has been influential [36] though contested [27]. Notable examples include comparing political decision-making in parliamentary and presidential systems. It is also relevant to the AI community and can shed some novel, interesting light on our discussion.

The debate in the USA over the militarisation of AI and LAWS can be viewed as the interaction of three actors: employees, management of AI organisations (academic groups or technology companies) and the government [38]. Crucially, one can view this as a veto player situation. The current status quo is one in which the AI community does not generally research LAWS. There are few US AI organisations, especially the most prominent, engaged in LAWS R&D for the US government. The US government, and management at several organisations, would like that to change – to a situation in which there is widespread collaboration from many organisations, including the most prominent. For the status quo to change, all actors must agree. However, a significant group of employees is not agreeing to this change.

In general, when are veto players more likely to be able to prevent a change of the status quo? Tsebelis [57] identifies three key factors:
1) the number of veto players;
2) lack of congruence; and
3) cohesion.

We suggest that these factors currently apply to some extent to the AI community, and are likely to largely continue to apply.
1) The number of veto players is not particularly high. It seems that perhaps management and the government believed they were in a two-player game. They seem not to have been prepared for the activism of the AI community on this topic, which indicates a three-player game.
2) The policy positions of the players have been rather dissimilar. Many governments and company managers want collaboration on military AI, and specifically LAWS, to not only be widespread but unproblematised. Many, perhaps most, employees do not.

This seems likely to continue, though, for example, the process of developing and exploring the proposed Department of Defense ethics principles [15] may achieve an accommodation between Silicon Valley and the Pentagon, increasing congruence between the players.
3) The relative similarity of policy positions amongst the AI community has been discussed above. There is clearly a debate about the extent to which they should be involved in the militarisation of AI, especially LAWS R&D, and whether there should be an international ban. This debate is likely to continue.

The AI community has achieved some successes as a veto player as the congruence between the players has been low (2), and the AI community has been fairly coherent (3). This situation is broadly likely to continue. However, whether the AI community will continue to be cohesive is a key question.

## ACKNOWLEDGMENTS

The author thanks Jess Whittlestone, Stephen Cave, Luke Kemp and Matthijs Maas for helpful comments and AI community activists and diplomats for anonymous interviews.